\documentclass[prl,twocolumn,showpacs,amsmath,amssymb,superscriptaddress]{revtex4-2}
\usepackage{graphicx}
\usepackage{hyperref}
\usepackage{xcolor}
\usepackage{color}
\usepackage[all,cmtip]{xy}
\usepackage{ulem}  %This package allows to strike text (tacharlo) with the command \sout 
\usepackage{amsmath,amsthm} 
\usepackage{amsfonts}

\makeatother

\begin{document}

%\title{Spin Hall transport in quasi nodal spheres}
%\title{On quasi-nodal spheres and the spin Hall effect}
%\title{\textcolor{red}{Quasi-nodal spheres: band inversion and spin Hall effect}\\
\title{On quasi-nodal spheres and the spin Hall effect: the case of YH$_3$ and CaTe.}
\author{Rafael Gonz\'{a}lez-Hern\'{a}ndez}
\email{rhernandezj@uninorte.edu.co}
\affiliation{Departamento de F\'{i}sica y Geociencias, Universidad del Norte, Km. 5 V\'{i}a Antigua Puerto Colombia, Barranquilla 080020, Colombia}
\affiliation{Institut f\"ur Physik, Johannes Gutenberg Universit\"at Mainz, D-55099 Mainz, Germany}
\author{Carlos Pinilla}
\email{ccpinilla@uninorte.edu.co}
\affiliation{Departamento de F\'{i}sica y Geociencias, Universidad del Norte, Km. 5 V\'{i}a Antigua Puerto Colombia, Barranquilla 080020, Colombia}
\affiliation{School of Chemistry, University of Bristol, Cantock's Close, Bristol, BS8 1TS, United Kingdom}
\author{Bernardo Uribe}
\email{bjongbloed@uninorte.edu.co, uribe@mpim-bonn.mpg.de }
\affiliation{Departamento de Matem\'{a}ticas y Estad\'{i}stica, Universidad del Norte, Km. 5 V\'{i}a Antigua Puerto Colombia, Barranquilla 080020, Colombia}
 \affiliation{Max Planck Institut fuer Mathematik, Vivatsgasse 7, 53115, Bonn, Germany}

\date{\today}

\begin{abstract}

Band inversion is a known feature in a wide range of topological insulators characterized by a change
of orbital type around a high symmetry point close to the Fermi level.  
In some cases of band inversion in topological insulators, the existence of quasi-nodal spheres has been detected, and
 the change of orbital type is  shown to be concentrated along these spheres in momentum space. 
 In order to understand this
phenomenon, we develop a local effective four-fold Hamiltonian which models the band inversion and reproduces
the quasi-nodal sphere. 
 This model shows  that 
 the signal of the spin Hall conductivity, as well as the change of orbital type, are both localized on the quasi-nodal sphere,
 and moreover, that these two indicators
 characterize the topological nature of the material.
 Using K-theoretical methods 
we show that  the  change of orbital type parametrized by an odd clutching function
 is equivalent to the strong Fu-Kane-Mele invariant.
We corroborate these results with \textit{ab-initio} calculations for the materials YH$_3$ and CaTe where
in both cases the signal of the spin Hall conductivity is localized on the quasi-nodal spheres in momentum space.
We conclude that a non-trivial spin Hall conductivity localized 
on the points of change of orbital type is a good indicator for Topological Insulation.

\end{abstract}
\maketitle

%\twocolumngrid
\section{Introduction}

%Borrador-por favor cambiar/quitar: 

The incursion of topological invariants in condensed matter physics has led to 
an enhanced classification of quantum materials \cite{topological-effects,classification-of-ti,RevModPhys.93.025002,RevModPhys.88.021004}. Among the insulators, the topological ones (TI) have attracted 
significant attention due to the presence of conducting surface states and because they show efficient spin transport properties \cite{qshe-graphene,QuantumSpinHallEffect-Bernevig,qshe-in-ti,SpinHallEffects-Sinova}. The presence of these unique surface states, together with the topological Fu-Kane-Mele invariant,  have been used to characterize these topological insulator phases from normal insulators systems \cite{Fu-Kane-Mele,Computing-topological-invariants,Kane-mele}.

The topological order of some insulators may be alternatively deduced from the change of orbital type
on the last valence bands around a time-reversal invariant point (TRIM). This feature is known in the literature as ``Band inversion'' and has been extensively used to classify TIs \cite{Colloquium-Topological-band-theory,PhysRevB.85.235401,ti-materials}.
Band inversion is assumed to be induced from band splitting due to strong spin-orbit coupling (SOC) interaction from heavy elements in materials \cite{Rashba-soc,soc-materials,soc-band-inv,Bi2Se3-ti,Strong-SOC}. 
In these systems, SOC can have a significant impact on the band structure and induces an opening gap between the conduction and valence bands thus changing the orbital type. 
This band inversion
may be coupled with
a change in the inversion symmetry eigenvalues at the appropriate TRIM. It was this precise feature that led to the establishment of the Fu-Kane-Mele invariant in topological insulators with inversion symmetry as Bi$_2$Se$_3$, Bi$_2$Te$_3$ and Sb$_2$Te$_3$ \cite{band-inversion-in-Bi2Se3,Bi2Se3-Bi2Te3-Sb2Te3}.
However, some materials show that the band inversion can appear spontaneously, or may be induced by strain, even without SOC interaction \cite{Strain-band-inversion-Sb,PhysRevB.85.235401}.
This phenomenon can be observed in $\alpha$-Sn, HgTe, several half-Heusler and chalcopyrite semiconductors, where the band inversion is presented at a particular TRIM and independent of the SOC interaction \cite{half-heusler-bands,ternary-heusler,Half-Heusler-Compounds,QSHE-soc-wells,chalcopyrite}

The type of Band inversion which is of interest for this work appears whenever there is a hybridization of the energy bands, and
the change of orbital type is concentrated on the ${\bf k}$-states where the energy gap is opened.
This form of band inversion is present, among others, on the trihydrides materials and it has been previously reported in references
\cite{nodal-line_YH3, PseudoDirac-nodal-sphere}.  On these materials a small energy gap is present and
the ${\bf k}$-states with a small energy gap form a 2-dimensional sphere.  In \cite{PseudoDirac-nodal-sphere}
these spheres were called ``Pseudo Dirac nodal spheres'' (PDNS) and here we have denoted  them ``Quasi-nodal spheres''.
In the pioneer work by \citet{PseudoDirac-nodal-sphere},  the possible formation of PDNS in crystal structures is reported on different symmetry point groups through the study of band crossings with pairs of 1D irreducible representations.  It is also found that the PDNS phase is robust against the SOC effect, in particular for the $M$H$_3$(with $M$=Y,Ho,Tb, Nd) and Si$_3$N$_2$ materials. 	In addition, the prediction of a realizable topological state with exotic transport properties is suggested for these PDNS prototypes \cite{PseudoDirac-nodal-sphere}. The analysis of the topological nature and the spin transport properties of the PDNS (or quasi-nodal spheres) remained an open question and it has been one of the motivations of the present work.
	
In this work, we study the mechanism underlying the band inversion occurring through hybridization
as it happens on the trihydrides.
We argue that the change of orbital type is concentrated on the quasi-nodal sphere, and moreover,
that the Fu-Kane-Mele invariant characterizing the TI property of the material can be deduced
from the non-triviality of the spin Hall conductivity (SHC) and the localization of its signal on the quasi-nodal sphere.
This fact explains the 2-dimensional 
nature of the ${\bf k}$-points where hybridization is happening and shows how the Fu-Kane-Mele invariant
relates to the SHC and the change of orbital type.

The paper contains two main parts, the theoretical analysis and the material realization.
In the first part, we analyze the mechanism of band inversion by using a four-fold effective Hamiltonian for a system where the orbital character 
change is present, the quasi-nodal sphere is induced, and both the signal of the SHC and the orbital type change are concentrated
on the quasi-nodal sphere.
 Here we show explicitly how the valence states of the effective Hamiltonian define an element
in the appropriate K-theory, and how this K-theoretical element matches the K-theoretical version of the strong Fu-Kane-Mele invariant.
In the second part, we investigate two possible materials, yttrium trihydride (YH$_3$) and calcium telluride (CaTe), which show an efficient spin/change conversion generated from the band inversion on the quasi-nodal sphere.

\section{Band inversion}

Let us consider a Hamiltonian with time reversal symmetry $ \mathbb{T}$ and inversion symmetry $\mathcal{I}$
where spin orbit coupling is taken into account. Band inversion could be understood as the change of orbital 
character on the last valence bands around a time reversal invariant point (TRIM).
This change of character,  
whenever coupled with a change of eigenvalue of the inversion operator, 
determines the strong topological insulator nature of the material.

In the case of interest, 
the change of orbital type is occuring on a sphere of points where the energy gap is small.
This 2-dimensional sphere in momentum space
with small energy gap (comparable to room temperature $\sim$ 25 meV) has been coined {\bf Quasi-nodal Sphere},
generalizing the concept of quasi-nodal line that was presented by the first and third authors in \cite{Gonzalez-Tuiran-Uribe-2021-quasi-nodal}.

In what follows we argue that band inversion due to hybridization produces a quasi-nodal sphere on momentum space where orbital characters 
 are mixed,  and moreover, where the Fu-Kane-Mele invariant can be extracted. 
 We first define a local model for
 the low energy effective Hamiltonian
inducing band inversion due to hybridization, we calculate its quasi-nodal sphere together with a $2 \times 2$ matrix information on each point obtained from the 
change of orbital type on the valence bands,
  and finally, we relate these parametrized matrices to the topological invariants obtained through K-theory.

\subsection{Model Hamiltonian}

Take the Pauli matrices $\tau_i$ and $\sigma_j$ in orbital and spin coordinates respectively, and
consider the Hamiltonian 
\begin{equation}  \label{model_hamiltonian}
H({\bf k}) = M({\bf k}) \tau_3 \sigma_0 + A({\bf k}) \tau_1 \sigma_3 + B({\bf k}) \tau_2 \sigma_0
\end{equation}
where 
\begin{align}
M({\bf k}) & = D_1 - m_1k_z^2 -n_1(k_x^2+k_y^2), \\
A({\bf k}) &  = D_2k_z +E_2k_z^3+ F_2(3k_x^2k_y-k_y^3), \\
B({\bf k}) & = D_3(k_x^3 -k_xk_y^2).
\end{align}

This Hamiltonian preserves time reversal, inversion and a three-fold rotation with matrices
\begin{align} \label{symmetries of Hamiltonian}
\mathbb{T}=i \tau_0\sigma_2 \mathbb{K},\ \ \ \mathcal{I}=- \tau_3\sigma_0, \ \ \mbox{and} \ \ C_3=e^{i \tau_0\sigma_3 \frac{\pi}{3}}
\end{align}
respectively. Here $C_3$ is a three-fold rotation around the $z$-axis. The bands are double-degenerate due to the presence
of both time reversal and inversion, and the energies of the bands are:
\begin{align}
\pm E({\bf k})= \pm \sqrt{M({\bf k})^2+A({\bf k})^2+B({\bf k})^2}.
\end{align}

\begin{figure*}
	\includegraphics[width=\textwidth]{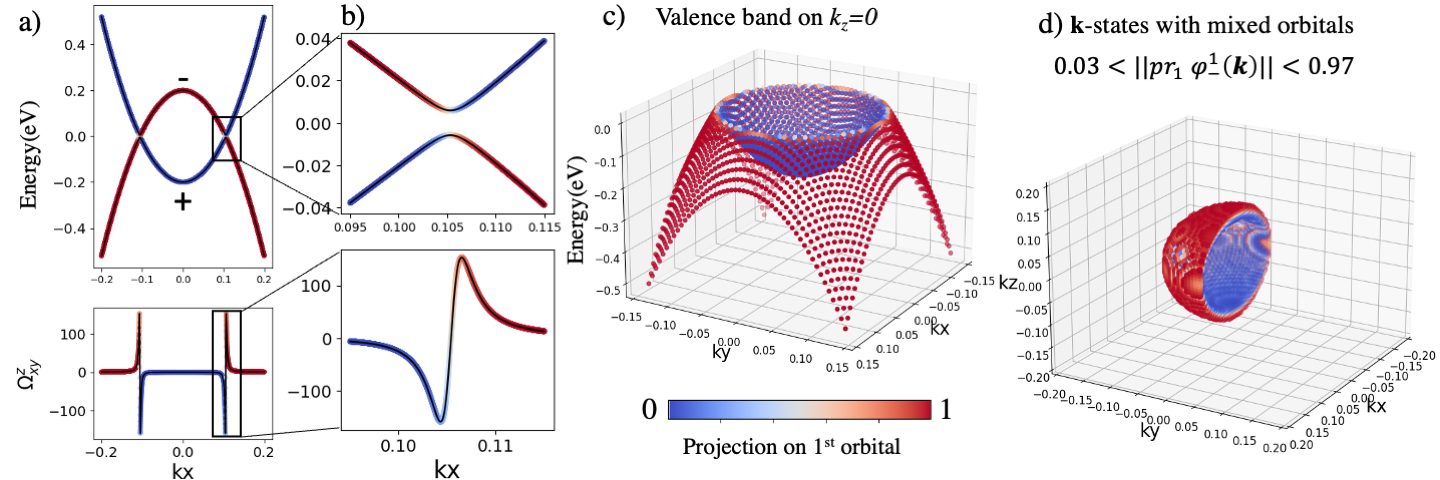}
	\caption{Energy bands of the Hamiltonian of Eqn. \eqref{model_hamiltonian} with projection to the first orbital.
a). Top panel: valence and conduction bands on $k_x$ coordinate axis with projection on the first orbital.
 The eigenvalues of the inversion operator at $(0,0,0)$ on the valence bands are $+1$ and on the
conduction bands are $-1$.  Bottom panel:
 sum of the spin Berry curvatures of the two valence bands $\Omega^z_{xy}$. 
 b) Top panel: close up view of the energy bands where 
the change of orbital type is concentrated.
 Bottom panel: 
 close up view of the sum of the spin Berry curvatures $\Omega^z_{xy}$.
 The signal of the 
 the spin Berry curvature
is concentrated where the change of orbital type occurs.
c) Valence band
on the plane $k_z=0$ with projection on the first orbital. 
d) ${\bf k}$-states with mixed orbital type (more than 3\% of each orbital type, $k_y \leq 0$) equivalent to the quasi-nodal sphere with energy gap of less than $0.05$eV. The color represents the magnitude of the projection on the first orbital. Inside the quasi-nodal sphere the states are concentrated on the first orbital, while outside are concentrated
on the second orbital. }
	\label{Fig_Hamiltonian}
\end{figure*}

Tuning up the constants to locally model the band inversion present in YH$_3$ at the point $\Gamma$, we set 
the coefficients to the following values:
\begin{align}\nonumber
D_1=&0.2, m_1=n_1=18,\\ D_2=&0.1, E_2=F_2=D_3=5.   \label{parameters}
\end{align} The energy bands of the Hamiltonian
are presented in Fig. \ref{Fig_Hamiltonian} a) and b) together with their projections on the first orbital.
There is no symmetry protecting energy crossings and therefore the eigenstates hybridize  producing an energy gap.
The change of orbital type is concentrated along the points in momentum space whose energy gap is small, and they
define a 2-dimensional sphere as can be seen in Fig. \ref{Fig_Hamiltonian} d).
 This 2-dimensional sphere is a quasi-nodal sphere and it is necessary for the band inversion
that induces the Hamiltonian. 
Note that the inversion operator $\mathcal{I}=- \tau_3\sigma_0$
is coupled with the orbital types since it acts by $-1$  on the first orbital and by $+1$  on the second orbital (see Fig. \ref{Fig_Hamiltonian} a).
The interdependence of the inversion operator with the orbit type is what permits
to distinguish the TI nature of the material by the change of eigenvalues of the inversion.
Nevertheless, the TI nature of the material is kept while inversion is broken, thus
the change of orbital type, together with the SHC is what allows to distinguish its topological nature.
 From Fig. \ref{Fig_Hamiltonian} d)
we see that the valence bands are concentrated on the second orbital inside the quasi-nodal sphere while
concentrated on the first orbital outside of it.

The change of orbital type of the valence bands along the quasi-nodal sphere permits to define a $2\times 2$
unitary complex matrix for each point on the nodal-sphere. This assignment produces a map from the sphere
$S^2$ to the Lie group of special unitary matrices $SU(2)$ which is equivariant with respect to the inversion operator. Here
the inversion operator acts by 
the antipodal action on the sphere and by multiplication by $-1$ on $SU(2)$. This map
can be understood as the clutching map (in the words of \citet[pp.20]{Atiyah}) that defines 
the rank $2$ complex vector bundle over the 3-dimensional sphere which produces the non-trivial Fu-Kane-Mele invariant.
Let us postpone the construction of the Fu-Kane-Mele invariant on the 3-dimensional sphere to the next section and
let us show how the matrices are defined.

Denote by $\varphi^{i}_\pm({\bf k})$, $i \in \{1,2\}$, the four eigenvectors of the Hamiltonian of Eqn. \eqref{model_hamiltonian}
with the fixed parameters described in Eqns. \eqref{parameters} satisfying the equation:
\begin{align}
H({\bf k})\varphi^{i}_\pm({\bf k}) = \pm E({\bf k}) \varphi^{i}_\pm({\bf k}).
\end{align}
Note that  the valence states $\varphi^{i}_{-}({\bf k})$ are well defined for all ${\bf k}$ 
since the system is gapped, and therefore we may take the vector projections of each valence state to the first and second orbital
respectively. The eigenstates have four coordinates
\begin{align}
\varphi^{i}_{-} = ( \varphi^{i}_{-,1}, \varphi^{i}_{-,2},\varphi^{i}_{-,3},\varphi^{i}_{-,4}) \in \mathbb{C}^4
\end{align}
and the projection on the first orbital takes the first two coordinates and the projection on the second takes the last two. Denoting
$pr_j$ the projection on the $j$-th orbital we have:
\begin{align}
pr_1 \varphi^{i}_{-}  = (\varphi^{ i}_{-,1}, \varphi^{i}_{-,2}),\\
pr_2 \varphi^{i}_{-}  = (\varphi^{ i}_{-,3}, \varphi^{i}_{-,4}).
\end{align} 

Along the quasi-nodal sphere both projections are non-trivial (we may take the sphere of radius $||{\bf k}||=0.11$, see Fig. \eqref{model_hamiltonian} c)), 
and therefore we may define the following matrix coefficients: 
\begin{align} \label{matrix coefficients}
A_{ij}({\bf k})= (-1)^{i} \left\langle \frac{ pr_1 \varphi^{i}_{-}({\bf k})}{||pr_1 \varphi^{i}_{-}({\bf k})||} \Bigg|  \frac{pr_2 \varphi^{ j}_{-}({\bf k})}{|| pr_2 \varphi^{ j}_{-}({\bf k}) ||} \right\rangle
\end{align}
for $i,j \in \{ 1,2 \}.$
Since the composition $\mathbb{T} \mathcal{I}$ commutes with the Hamiltonian, we may consider $\varphi^{2}_{-}$
to be the Kramer's pair of $\varphi^{1}_{-} $ satisfying the equation 
\begin{align} \label{Kramer pair}
\varphi^{2}_{-}=\mathbb{T} \mathcal{I}\varphi^{1}_{-}.
\end{align}
From Eqns. \eqref{symmetries of Hamiltonian} we know that $\mathbb{T} \mathcal{I}=-i\tau_3\sigma_2 \mathbb{K}$ and therefore 
Eqn. \eqref{Kramer pair} implies the following equality:
\begin{align} \label{Kramer pair coordinates}
 ( \varphi^{2}_{-,1}, \varphi^{2}_{-,2},\varphi^{2}_{-,3},\varphi^{2}_{-,4}) = (-\overline{\varphi}^{1}_{-,2}, \overline{\varphi}^{1}_{-,1},-\overline{\varphi}^{1}_{-,4},\overline{\varphi}^{1}_{-,3}).
\end{align}

Replacing Eqn. \eqref{Kramer pair coordinates} on the definition of the matrix coefficients of Eqn.  \eqref{matrix coefficients} we 
see that the matrix $A$ is unitary with $A_{11}=\overline{A}_{22}$, $ A_{12}=- \overline{A}_{21}$, and  its determinant
is $1$. Hence the matrix $A$ belongs to the Lie group $SU(2)$ of special unitary matrices.

If we take $S^2_r=\{{\bf k} \colon ||{\bf k}||=r\}$ to be the quasi-nodal sphere of the system ($r=0.11$ for the Hamiltonian 
with parameters in Eqns. \eqref{parameters}), we obtain a map
 \begin{align}
A : S^2_r \to SU(2), \ \ \ {\bf k } \mapsto A({\bf k}).
\end{align}

The explicit form of the inversion operator  $\mathcal{I}=- \tau_3\sigma_0$ implies the following equation:
\begin{align}
pr_j \varphi^{i}_{-}({\bf k}) = (-1)^j pr_j \varphi^{i}_{-}(-{\bf k}), \ j \in \{1,2\},
\end{align}
which makes the map $A$ equivariant with respect to the inversion action, i.e.
\begin{align} \label{equivariant A}
A(-{\bf k}) = -A({\bf k}).
\end{align}

Any equivariant map such as $A$ produces the Fu-Kane-Mele invariant since any two maps
satisfying Eqn. \eqref{equivariant A} are homotopic \cite[Lem. 3.27]{Gomi}. In particular the map $A$ is homotopic
to the map $ C: S^2_r \to SU(2)$,
\begin{align} \label{clutching function}
   C({\bf k })=
\left( \begin{matrix}
ik_z & k_x+ik_y \\
-k_x+ik_y & -ik_z,
\end{matrix} \right),
\end{align}
which is a simple clutching function that defines the rank $2$ bundle with non-trivial Fu-Kane-Mele invariant.

Another important feature of the quasi-nodal sphere of this Hamiltonian is the fact that the signal
 for the spin Hall conductivity $\Omega^z_{xy}$ localizes around it. 
In Fig. \ref{Fig_Hamiltonian} d) we have plotted the points on which the orbital type is mixed, i.e.
${\bf k}$-points with more than $3\%$ on each orbital
\begin{align}
0.03 < ||pr_1 \varphi^{1}_{-}({\bf k})||<0.97,
\end{align} 
and we notice that these mixed states define the quasi-nodal sphere.  From Fig. \ref{Fig_Hamiltonian} b)
we notice that the signal 
of the SHC
 is localized on the ${\bf k}$-points with mixed orbital type. 
This feature could be appreciated on the materials YH$_3$ and CaTe on  Fig. \ref{Fig1} d), f), g) and Fig. \ref{Fig2} d), f), g) respectively.

The topological invariant that the Hamiltonian of Eqn. \eqref{model_hamiltonian} defines can be better understood with the help
of K-theory. This is the subject of the next section.

\subsection{K-theory, clutching functions and sewing matrices}

Restrict the Hamiltonian to a ball $B$ centered at the origin which includes the quasi-nodal sphere. The valence
eigenstates define a rank $2$ complex vector bundle over $B$ which incorporates the clutching function
$A$. This bundle possesses the topological information to capture the Fu-Kane-Mele invariant, but unfortunately
it does not trivialize on the boundary of $B$ in order to define an appropriate element in K-theory (the Hamiltonian of Eqn. \eqref{model_hamiltonian} provided
only a local model and it was not defined over a compact space). To overcome this issue
 we will construct a rank $2$ complex vector bundle
over the sphere $S^3$ which models the structure of the valence eigenstates of the Hamiltonian of Eqn. \eqref{model_hamiltonian}
and we will show that this vector bundle incorporates the Fu-Kane-Mele invariant for K-theory.
This bundle will model the K-theoretical properties of the valence states, and will permit us to calculate its topological invariants.

Consider the 3-dimensional sphere
\begin{align}
S^3=\{(t,k_1,k_2,k_3) | t^2+||k||^2=1 \},
\end{align}
where $S^3 \backslash \{(-1,{\bf 0})\}$ could be thought as the stererographic projection of the ball $B$ and $S^3$ its
one-point compactification. Take
 $\rho : [-1,1] \to [0,1]$ a partition of unity on the interval $[-1,1]$ with $\rho|_{[-1,-1+\varepsilon)}=0$  and $\rho|_{(1-\varepsilon,1]}=1$ for small $\varepsilon$. Define the rank $2$ vector bundle $E \subset S^3 \times (\mathbb{C}^2)^2$
 by the equation
 \begin{align} \label{E_on_C4}
E = \left\{ \left[(t,{\bf k}),   (\rho(t) u, (1-\rho(t)) C({\bf k}) u) \right]  \ | \ u \in \mathbb{C}^2  \right\},
 \end{align}
and define the actions of $\mathcal{I}$ and $\mathbb{T}$ as in Eqns. \eqref{symmetries of Hamiltonian}
 by the following formulas:
\begin{align}
\mathcal{I} \cdot \left((t,{\bf k}),(u,v) \right) &= \left((t,-{\bf k}),(-u,v) \right), \\
\mathbb{T} \cdot \left((t,{\bf k}),(u,v) \right) &= \left((t,-{\bf k}),(\mathbb{J}u,\mathbb{J}v) \right).
\end{align}
Here $\mathbb{J} = i \tau_2 \mathbb{K}$ commutes with the matrices in $SU(2)$ and therefore the action
of $\mathbb{T}$ is well-defined. Moreover,  the partition of unity $\rho$ could be understood as the 
projection map on the first orbital while $(1-\rho)$ as the projection on the second.

When the action of $\mathcal{I}$ is disregarded, the relative bundle $[E] -[S^3 \times \mathbb{C}^2]$
generates the only non-trivial class on the appropriate relative K-theory groups. Choosing as base
point $(-1, {\bf 0})$ and following the notation of
\cite[App. C]{Gomi}, we have that
\begin{align}
[E] - [S^3 \times \mathbb{C}^2]  \in \widetilde{KQ}^0(S^3) \cong \mathbb{Z}/2,
\end{align}
and therefore the bundle $E$ induces the strong Fu-Kane-Mele invariant. Here $KQ^{n+4} \cong KR^n$ where $KR^*$ is Atiyah's real K-theory \cite{Atiyah-real}, and 
\begin{align}
\widetilde{KQ}^0(S^3) \cong \widetilde{KR}^4(S^3) \cong KR^7(*) \cong KO^{-1} \cong \mathbb{Z}/2.
\end{align}

Taking into account the involution
$\mathcal{I}$, the relative bundle $ [E] - [S^3 \times \mathbb{C}^2]$ gives the generator of the relative
equivariant symplectic K-theory groups:
\begin{align}
[E] - [S^3 \times \mathbb{C}^2]  \in \widetilde{KSp}^0_{\langle \mathcal{ I} \rangle}(S^3) \cong \mathbb{Z}.
\end{align}
Here the composition $\mathcal{I}\mathbb{T}$ defines the quaternionic structure on the fibers of bundle
and $\mathcal{I}$ induces a $\mathbb{Z}/2$-equivariant quaternionic action on the bundle. The groups $\widetilde{KSp}^0_{\langle \mathcal{ I} \rangle}(S^3)$
are the reduced $\mathbb{Z}/2$-equivariant symplectic K-theory groups of the sphere $S^3$ \cite{Mishchenko}, cf. \cite{Dupont}.
The restriction map to the point $(1,{\bf 0})$ counts the number of quaternionic $\mathbb{Z}/2$-representations with $\pm 1$-eigenvalues:
\begin{align}
\widetilde{KSp}^0_{\langle \mathcal{ I} \rangle}(S^3) \to&   \mathbb{Z}_{-1}\oplus \mathbb{Z}_{+1} \cong
KSp^0_{\langle \mathcal{ I} \rangle}(\{(1,{\bf 0})\}),\\
E \mapsto &1 \oplus 0,
\end{align}
where  $\mathbb{Z}_{-1}$ counts the non-trivial ones and $ \mathbb{Z}_{+1}$ the trivial ones. 
Since this restriction map is injective, the topological class of the bundle can be determined by the
eigenvalues of $\mathcal{I}$ on the fixed point set of the action. Since $E$ has the non-trivial
quaternionic $\mathbb{Z}/2$-representation on the fixed point $(1,{\bf 0})$ and the trivial one on 
$(-1,{\bf 0})$, the strong Fu-Kane-Mele invariant can be determined by the parity of the number
of pairs of complex $-1$ eigenvalues on the fixed points of the $\mathcal{I}$ action; this is the main result of \citet{Fu-Kane-Mele}.

The restriction of the bundle $E$ to the sphere of points with $t=0$, permits us to the define the transformation
from the first to the second orbital, and this transofrmation is precisely the function $C$ presented in Eqn. \eqref{clutching function}.
This clutching function is shown in \cite[Cor. 4.1]{Gomi} to induce the Fu-Kane-Mele invariant.

The bundle $E$ models the topological structure of the valence states of the Hamiltonian of Eqn. \eqref{model_hamiltonian} and it
induces the topological generators in both the equivariant
symplectic K-theory and the quaternionic K-theory of involution spaces. Hence the bundle
$E$ models the topological nature of the band inversion and it induces the strong Fu-Kane-Mele invariant \cite{Fu-Kane-Mele}.
The change of eigenvalues of the inversion operator on the fixed points on the sphere $S^3$ is only possible whenever
the clutching function $C$ is defined on the sphere $S^2$ and it is odd (Eqn. \eqref{equivariant A}). 

\begin{figure}
	\includegraphics[width=8.5cm]{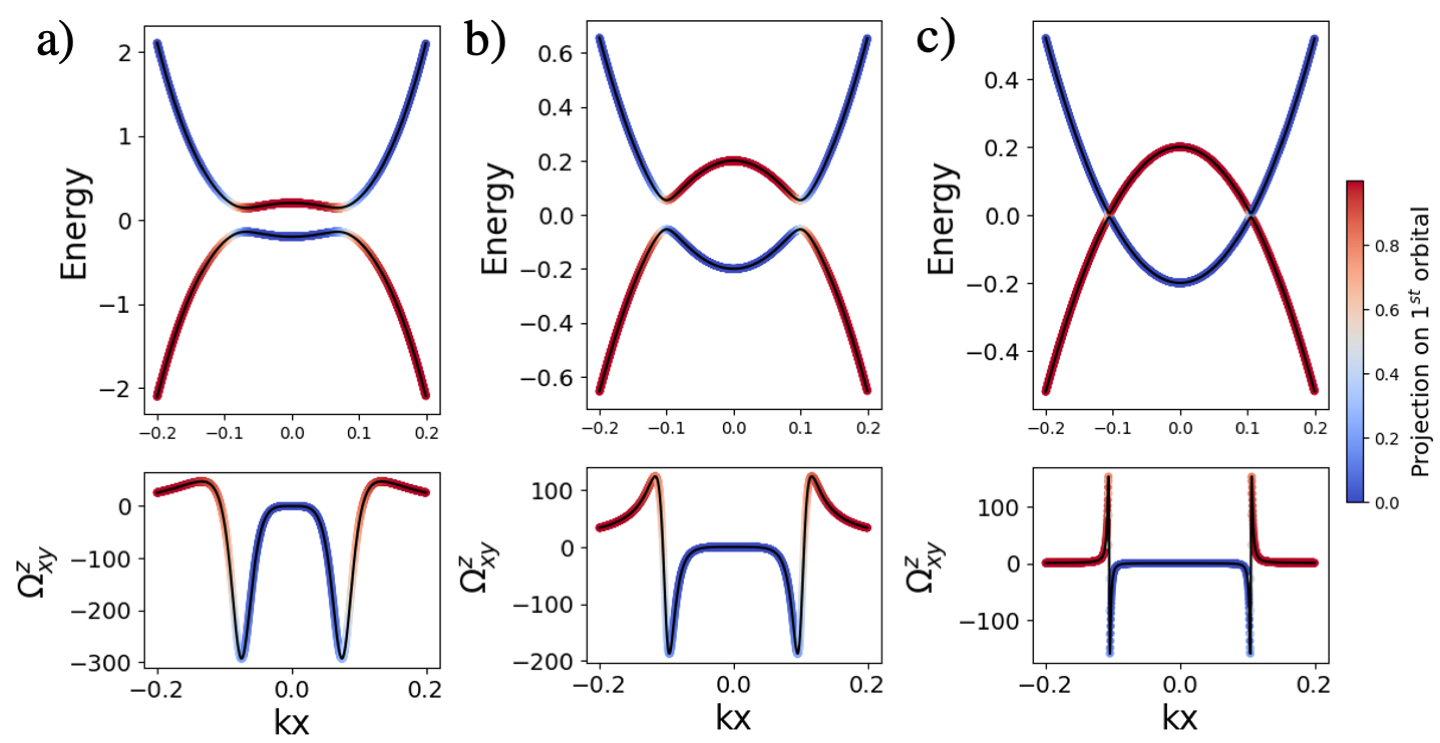}
	\caption{Spin Berry curvature $\Omega^z_{xy}$ at the Fermi level with respect to $k_x$
	of the Hamiltonian of Eqn. \eqref{model_hamiltonian}. The energy gap is widened by setting the structural constants $E_2,F_2,D_3$ of
    the Hamiltonian to the following values: a) $E_2=F_2=D_3=250$ b) $E_2=F_2=D_3=50 $ and c) $E_2=F_2=D_3=5$. The values
    of c) are the original values from Eqn. \eqref{parameters}.
	The signal of the SHC is concentrated on the region where the orbital type changes. Note that despite the fact that
	the difference of energies appears on the denominator of the Kubo formula of the SHC (see Eqn. \eqref{formula omegaz_xy}),
	the signal of the spin Berry curvature localizes whenever the change of orbit type occurs.    }
	\label{FigOmegaz_xy}
\end{figure}  
% =  \Omega^z_{1,xy}+\Omega^z_{2,xy}

The relation of the clutching functions with the SHC can be seen in Fig. \ref{FigOmegaz_xy}.
The signal of the SHC concentrates on the points where the change of orbital type occurs despite the size of the energy gap.
We can therefore conclude that not only the ${\bf k}$-points with mixed orbital type make the quasi-nodal sphere, but moreover,
that the signal of the spin Hall conductivity localize on this quasi-nodal sphere.

We can go one step further and we may calculate the sewing matrices
of the inversion and the time reversal operator. For this we need to take a trivialization of the bundle
$E$ (since any rank $2$ complex bundle over $S^3$ is trivializable), and we may write
the inversion and the time reversal operator on the new basis. Now we have $E = S^3 \times \mathbb{C}^2$
and the actions become
\begin{align} \label{action_I}
\mathcal{I} \cdot \left((t,{\bf k}),u \right) &= \left((t,-{\bf k}),(t\mathrm{Id} + C({\bf k}))u \right), \\
\mathbb{T} \cdot \left((t,{\bf k}),u \right) &= \left((t,-{\bf k}), (t\mathrm{Id} + C({\bf k}))\mathbb{J}u \right).
\label{action_T}
\end{align}

For $(e_1,e_2)$ the canonical basis on $\mathbb{C}^2$, the sewing matrices 
for the operator $P$ on $E$ are defined by the equation

\begin{align}
G(P)_{ij}(t, {\bf{k}})= \langle e_i | P(t, {\bf k}) | e_j \rangle.
\end{align}
For the inversion and the time
reversal operator the sewing matrices can be described as follows

\begin{align}
G(I)(t, {\bf k}) &= t\mathrm{Id} + C({\bf k}), \\
G(\mathbb{T})(t, {\bf k}) &= (t\mathrm{Id} + C({\bf k}))\mathbb{J}.
\end{align}
In both cases the sewing matrices define maps from $S^3$ to $SU(2)$ whose degree is $\pm1$. From
\cite{Symmetry_representation_approach_to_topological_invariants, 
Equivalent_topological_invariants_of_topological_insulators,
Inversion_symmetric_topological_insulators}
we know that the parity of the degree of those maps recovers the Chern-Simons axion coupling term

\begin{align}
\theta = \frac{1}{4\pi} \int_{\mathrm{BZ}} dk^3 \epsilon^{\alpha \beta \gamma} \mathrm{Tr} \left( \mathcal{A}_\alpha
\partial_\beta \mathcal{A}_\gamma - \frac{2i}{3} \mathcal{A}_\alpha \mathcal{A}_\beta \mathcal{A}_\gamma \right).
\end{align}

Since the parity of the degrees of both $G(I)$ and $G(\mathbb{T})$ is odd, then we know that $\theta=\pi$.
Hence we see the topological nature of the trivial complex bundle $E$ whenever the action of either $I$ or
$\mathbb{T}$ are defined as in Eqns. \eqref{action_I}, \eqref{action_T}. 

Note that whenever $t=0$, both the clutching function of the bundle of Eqn. \eqref{E_on_C4} and the sewing matrices of the inversion operator agree. The topological nature of the bundle could be theoretically deduced then from either one. Unfortunately both approaches are not well suited for computational calculations. On the one hand the definition
of the clutching functions is not gauge invariant, and on the other, the calculation of the Chern-Simons coupling term
has been until now elusive for real materials \cite{Gauge-discontinuity_Chern-Simons}. We hope in the future new methods are developed to calculate both.

\section{Material realization}
\subsection{Yttrium trihydride}
\begin{figure*}
	\includegraphics[width=17.4cm]{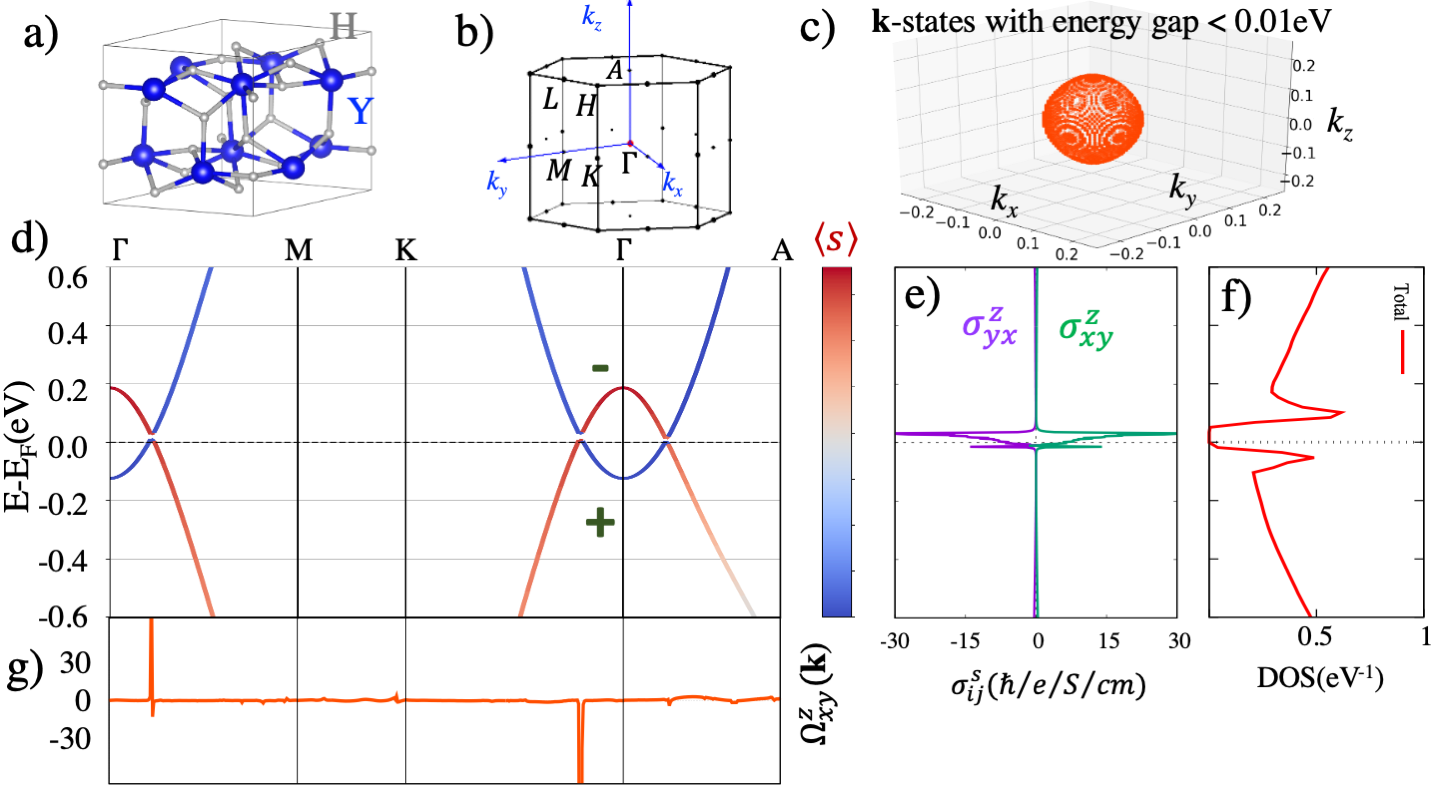}
	\caption{a) Ball stick model of the hexagonal crystal structure of YH$_3$ material \cite{vesta}. b) Hexagonal Brillouin zone indicating the high-symmetry points. c)  Position of electronic states in reciprocal space with energy band gap of less than 0.01 eV. d) Projected band structure with H-$s$ orbitals along the high symmetry lines, where the symbols + and - indicate the inversion symmetry eigenvalues
	of  +1 and -1 respectively at $\Gamma$ point. e) Spin Hall conductivity (in $\hbar$/$e$/S/cm). f) Total density of states (DOS) as a function of Fermi energy. 
	g) Spin Berry curvature $\Omega^z_{xy}({\bf k})$ [$\r{A}^2$] along high symmetry lines added over the valence band. $\Omega^z_{xy}({\bf k})$ signal is concentrated where the change of orbital character occurs.
	  It is noted a large SHC in the band gap where there are no electronic states. Spin-orbit coupling is included in all the calculations and the Fermi level is set to zero.}
	\label{Fig1}
\end{figure*}  

The ytrium trihydride (YH$_3$) compound can crystallize in a hexagonal structure with space group P-3c1  \#165 \cite{yh3-p3m1}, as shown in Fig. \ref{Fig1} a). This space group contains 12 symmetry operations which can be generated by the rotation $C_{3z}$, the screw rotation $S_{2(x+y)}$, the inversion $\mathcal{I}$ and the time-reversal $\mathbb{T}$ symmetry operations. The presence of both $\mathcal{I}$ and $\mathbb{T}$ symmetries indicate that energy bands
 are double degenerate along the full Brillouin zone (BZ). The BZ for this material is an hexagonal unit cell in  reciprocal space with high-symmetry points as indicated in Fig. \ref{Fig1} b).

Based on \textit{ab-initio} calculations, we have obtained the electronic band structure
 of the YH$_3$ material and we have confirmed the appearance of the quasi-nodal 
 sphere centered at the $\Gamma$ point. The band structure along the high-symmetry
  lines, including the spin-orbit coupling  interaction, is shown in Fig. \ref{Fig1} d). 
  It should be noted
 a hybridization band gap between the valence and conduction bands for all 
 $\bf k$-paths that connect the  $\Gamma$ point with any other high symmetry point. 
 The hybridization band gap is originated by the SOC interaction and it is extended
 along the quasi-nodal sphere in  reciprocal space. Symmetry analysis indicates that no
  crystal symmetry protects a possible nodal sphere or line whenever SOC is included. 
 To confirm the existence of the quasi-nodal sphere, we carried out a 
 systematic search of the $\bf k$-points in BZ where almost zero band gaps
 between the occupied and unoccupied bands are located. To visualize the 
 quasi-nodal sphere, we have plotted the $\bf k$-points in the BZ with energy 
  band gap of less than $0.01$ eV in Fig. \ref{Fig1} c).

The orbital-resolved H-$s$ projection is also shown in the band structure of Fig. \ref{Fig1} d). The figure shows that these orbitals dominate the valance band close to the Fermi level. In addition, a band inversion (change of the orbital character) is noted around the  $\Gamma$ point, which is occurring on the surface of the quasi-nodal sphere, as it was predicted by the $\bf k \cdot p$ model presented in the previous section.

We have also calculated the inversion symmetry eigenvalues on all TRIMs and we have
 observed the change of eigenvalues from the valence to the conduction bands at the 
 $\Gamma$ point (symbols + and - in Fig. \ref{Fig1} d). This change of eigenvalues
 indicates that the material can be considered a topological insulator with a tiny
 band gap spread along the quasi-nodal sphere. We have corroborated the topological insulator property of this material by using the Wilson loop method in order to find the
$\mathbb{Z}_2$ invariants for the six $k_i$ planes in the BZ. Our calculations show that
for the $k_i$=0 planes the invariant is $1$ while for the $k_i$=$ \pi$ planes the invariant is
$-1$ (here $i$=$x,y,z$). These results indicate the existence of the strong 
$\mathbb{Z}_2$ Fu-Kanel-Mele topological index in the YH$_3$ material.

As it is well known, topological insulators are materials that can provide a platform to reach a large spin Hall
 conductivity (SHC) signal \cite{qshe-in-ti,SpinHallEffects-Sinova}. For the case of YH$_3$, we have plotted the SHC as a function of Fermi level
  in Fig. \ref{Fig1} e). 
   In Fig. \ref{Fig1} g)  we found a strong SHC response at the Fermi level,
   well distributed all around the
   quasi-nodal sphere. 
  Note that the shape of the signal of the SHC along $\Gamma$-M matches the one of the model Hamiltonian presented
   in Fig. \ref{Fig_Hamiltonian}.  a)
 This particular feature has been also observed in 2D topological insulators, where the $\bf k$-resolved SHC signal is distributed around the $\Gamma$ point \cite{she-in-2d-ti}.
 Furthermore, it is important to highlight that a strong SHC signal is observed in the absence of energy states at the Fermi level, as it can be observed in the total density of states calculation in Fig. \ref{Fig1} f).  
 These results indicate that hydride materials can be promissory materials to efficient spin/charge conversion by means of the SHE \cite{she-dihydrides}. 

\begin{figure*}
	\includegraphics[width=17.8cm]{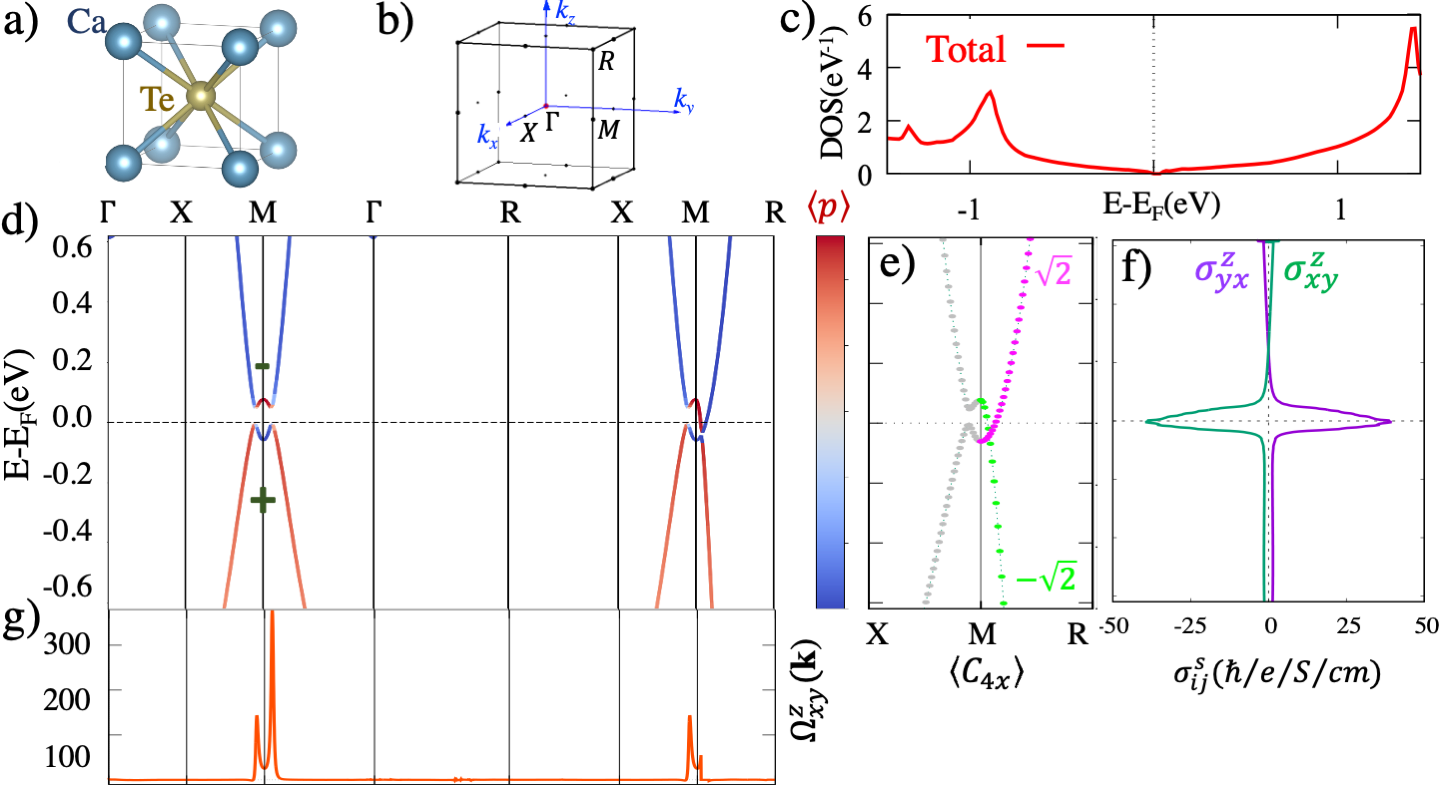}
	\caption{a) Crystal structure \cite{vesta} and b) cubic Brillouin zone indicating the high-symmetry points of the CaTe material (cubic phase). c)  Total density of states (DOS) as a function of Fermi energy. d) Projected band structure with Ca-$p$ orbitals along the high symmetry lines, where the symbols + and - denote the inversion symmetry eigenvalues at M point, +1 and -1, respectively. e)  $C_{4z}$ symmetry eigenvalues resolved at the electronic bands along the X-M-R $k$-path. f) Spin Hall conductivity (in $\hbar$/$e$/S/cm) as a function of Fermi energy.  It is noted a strong  SHC signal at the Fermi level, which is attributed to the quasi-nodal sphere and the Dirac point. g) Spin Berry curvature $\Omega^z_{xy}(k)$ [$\r{A}^2$] along high symmetry lines added over the valence band. $\Omega^z_{xy}(k)$ signal is concentrated where the change of orbital character occurs. Spin-orbit coupling is included in all the calculations and the Fermi level is set as zero.}
	\label{Fig2}
\end{figure*}  

\subsection{Calcium telluride}

Calcium telluride (CaTe) is another material with quasi-nodal sphere realization due to band inversion.
CaTe is a non-magnetic material whose space group is Pm-3m  \#221 and has 48 symmetry
operations. This symmetry group can be generated by the operations: $C_{2z}$, $C_{2y}$,
$C_{2(x+y)}$, $C_{3(x+y+z)}$, $\mathcal{I}$ and $\mathbb{T}$. The cubic crystal
structure and Brillouin zone for the CaTe compound are shown in Fig. \ref{Fig2} a) and
Fig. \ref{Fig2} b), respectively. In Fig. \ref{Fig2} d) we present the electronic band
structure with the orbital-projected Ca-$p$ states. We note a band character inversion
between the valence and conduction bands around the M point. The symbols + and - in
Fig. \ref{Fig2} d) indicate a reversal of the inversion symmetry eigenvalues at the 
M point, which indicates a topological response on this compound.

We have also found a Dirac point (DP) along the M-R ${\bf k}$-path that can be observed in the band structure of Fig. \ref{Fig2} e).
 However, a tiny band gap is observed along any ${\bf k}$-path starting from M when the SOC is taken into account.
  The stability of the DP is confirmed by the projection of the $C_{4x}$ rotation symmetry eigenvalues at the two
   bands that generate the DP as shown in Fig. \ref{Fig2} e). The eigenvalues of $C_{4x}$ are the fourth roots of $-1$ and by Kramer's
   rule they come in pairs: $\left(e^{i\frac{\pi}{2}},e^{-i\frac{\pi}{2}}\right)$ and $\left(e^{i\frac{3\pi}{}2},e^{-i\frac{3\pi}{2}}\right)$. The trace of the 
   associated matrix permits us to detect the type of corepresentation  at the high symmetry line and these traces are $\sqrt{2}$ and $-\sqrt{2}$ respectively. Whenever
   the traces differ, it means that the bands have different corepresentations of the group generated by $C_{4x}$ and $\mathcal{I}\mathbb{T}$
   and therefore the bands cannot hybridize \cite{Gonzalez-Tuiran-Uribe-2021-chiralities}. In this particular case the traces of the operator $C_{4x}$ are different 
  and therefore the DP is protected by the four-fold rotation symmetry.

The procedure for finding the quasi-nodal sphere was also applied in the CaTe case.
The quasi-nodal spheres are centered at the M points,
and the signal of the ${\bf k}$-resolved spin Berry curvature peaks at the
 ${\bf k}$-points where the quasi-nodal sphere is located as shown in Fig. \ref{Fig2} g). We  found that the SHC response of the quasi-nodal sphere arises from the band inversion around the M point, as illustrated in Fig. \ref{Fig_Hamiltonian} b) obtained from the ${\bf k \cdot p}$ Hamiltonian model of Eqn. \eqref{model_hamiltonian}. In the case of CaTe this feature is noticed, namely that the regions near the quasi-nodal sphere (M-point centered) retain locally large ${\bf k}$-resolved SHC (see Fig. \ref{Fig2} g)).

\begin{figure}
	\includegraphics[width=8.45cm]{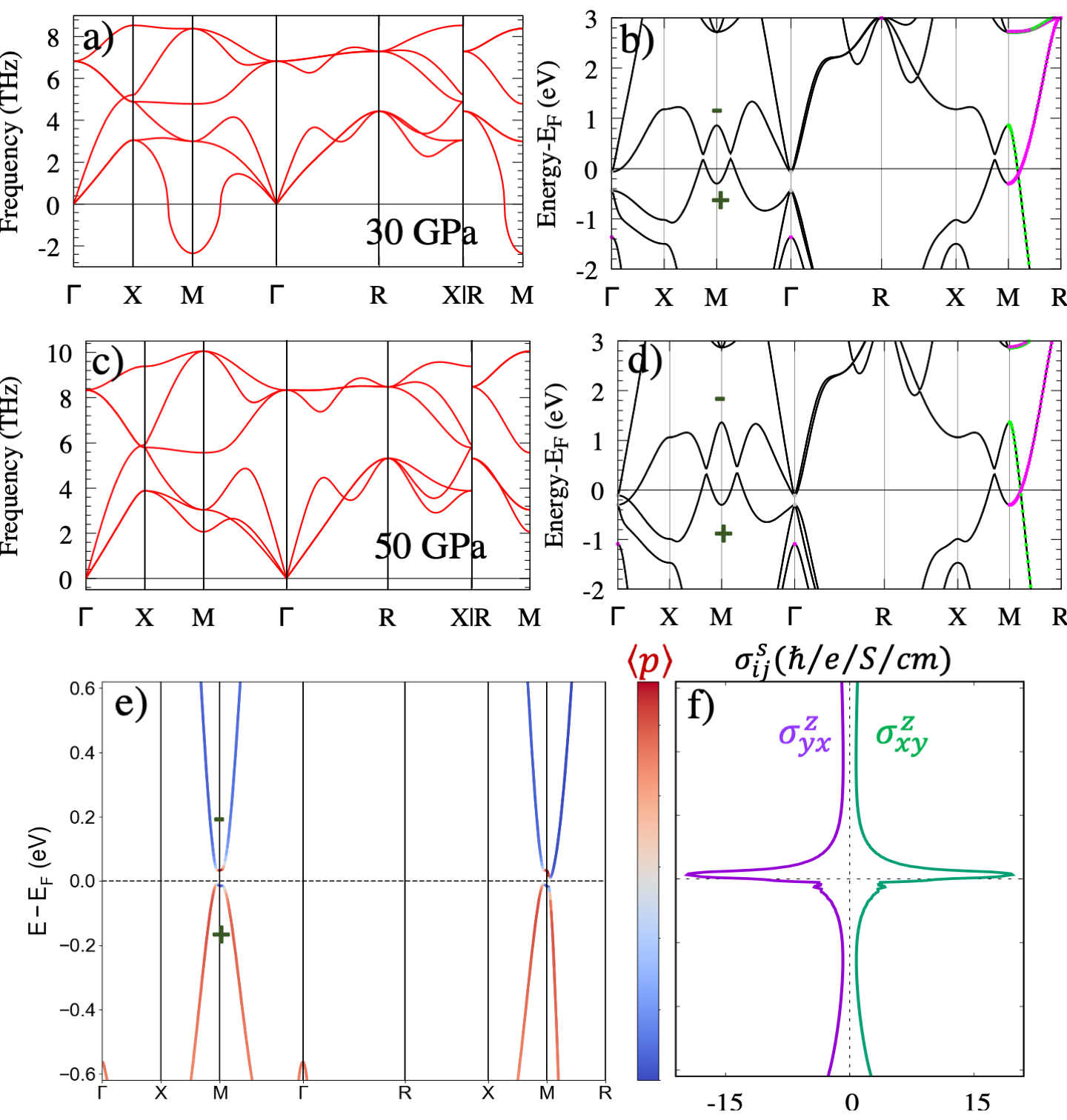}
	\caption{Top four panels: dynamical stability of cubic CaTe under high pressure. a) and c) correspond to phonon dispersion curves at 30 GPa and 50 GPa respectively. Electronic band structure at the same given pressures are contained in panels b) and d) with preserved band inversion at $\Gamma$ point. The colors indicate the eigenvalues of the four-fold rotation $C_{4x}$. Bottom two panels:
	CaTe with a small ($\epsilon$=3\%) diagonal distortion, thus breaking the four-fold symmetry. 
	e) Electronic band structure with gapless phase; note that there is no DP due to the broken four-fold symmetry.  
	 f) Energy resolved spin Hall conductivity.  The SHC arises mainly from the electronic bands that generate the quasi-nodal sphere and it survives despite the absence of the DP.}
	\label{Fig3}
\end{figure}  

Topological insulators share a strong spin Hall conductivity, however, the presence of a 
Dirac point is not always an indication of SHC. We have found a large signal of the SHC
at the Fermi level for the CaTe as shown in \ref{Fig2} f), where the spin Hall conductivity
as a function of Fermi energy is presented. We can notice that the SHC signal is strong
and coincides with the energy range of the quasi-nodal sphere and the DP.  However,
it is not clear whether the signal of the SHC comes from the quasi-nodal sphere,
the DP or both. In order to distinguish the contribution to the spin Hall conductivity 
by the DP and the quasi-nodal sphere we have applied a small (3\%) diagonal distortion 
on the cubic CaTe. This strain deformation transforms CaTe from the cubic (Pm-3m) to 
the rhombohedral (R-3m) structure. The deformation preserves the symmetry 
generators: $C_{2(x+y)}$, $C_{3(x+y+z)}$, $\mathcal{I}$ and $\mathbb{T}$, 
but breaks the $C_{4x}$ symmetry. The breaking of the $C_{4x}$ symmetry leads 
to a band gap and a subsequent absence
of the DP is noticed (see Fig. \ref{Fig3} e).

The rhombohedral structure of CaTe preserves the band inversion, the reversal of the inversion symmetry eigenvalues, and the
 quasi-nodal sphere around the M point. For this rhombohedral phase, it is noted a topological insulator behavior with a 
 small band gap, similar to the YH$_3$ case and the four-states Hamiltonian model of Eqn. \eqref{model_hamiltonian}.
 The calculation shows a non-zero SHC in the band gap, which emerges from the band inversion at the quasi-nodal
  sphere as shown in Fig. \ref{Fig3} f) and corroborated by the ${\bf k \cdot p}$ model (Fig. \ref{Fig_Hamiltonian} b). 

The CaTe crystalline phase displayed in Fig. \ref{Fig2} a) has been experimentally known to be dynamically stable for pressures above 33 GPa\cite{Luo1994}. This has also been corroborated by our $ab$-$initio$ calculations (see figures \ref{Fig3} a) and \ref{Fig3} c). We can see that the DP along the M-R {\bf k}-path is preserved even when an isotropic pressure is applied (see figures \ref{Fig3} b) and \ref{Fig3} d)). This result is expected since the $C_{4x}$ symmetry, band inversion and quasi-nodal sphere are not affected by an isotropic compression.

Our results suggest that CaTe  and
YH$_3$ could both display large spin Hall angles for practical applications since they show small longitudinal conductivities (low density of states) at the Fermi level. The SHC signal is concentrated at the hybridized band regions, as it was also found by \citet{she-due-hybridized}.

\subsection{Computational methods}

For YH$_3$ and CaTe materials, we have carried out $ab$-$initio$ calculations within the density-functional theory with an uniform {\bf k}-mesh in the Brillouin zone of 13$\times$13$\times$11 and  14$\times$14$\times$14, respectively. 
The energy cut-off was taken to be 520 eV and the exchange-correlation contribution to the total energy treated within the PBE parametrization \cite{Perdew1996}, as implemented in the Vienna \textit{ab-initio} simulation package (VASP) \cite{vasp}.  
Projected band structures were plotted using the {\tt pyprocar} program \cite{pyprocar}.
The symmetry eigenvalues along the high symmetry lines were calculated using the {\tt irrep} code \cite{irrep}. 
To estimate the spin transport properties, we employed the {\tt wannier90} code \cite{wannier90,pizzi_2020} to build a Wannier Hamiltonian.  The intrinsic spin Hall conductivity (SHC) was evaluated by integrating the spin Berry curvature on a dense 240$\times$240$\times$240 grid in the first Brillouin zone, using the {\tt linres} code \cite{SHEcode}. In this model, the SHC can be written as:
\begin{equation}
\sigma^z_{xy}=-\dfrac{e^2}{\hbar} \sum_{n}\int_{BZ}\dfrac{dk^3}{(2\pi)^3}f_n(k)\Omega^z_{n,xy}(k),
\end{equation}
where $f_n(k)$ is the Fermi-Dirac distribution and $\Omega^z_{n,xy}(k)$ is the spin Berry curvature for the $n$th band. The spin Berry curvature for the $n$th band can be calculated using the Kubo formula:
\begin{equation} \label{formula omegaz_xy}
\Omega^z_{n,xy}(k)= -2{\hbar}^2 \text{Im}\sum_{m\neq n}%
\frac{\langle n,k\vert \widehat j^z_{x} \vert m,k \rangle\langle m,k\vert \widehat v_y \vert n,k  \rangle}{\left(\epsilon_{n,k}-\epsilon_{m,k}\right)^{2}}, 	
\end{equation}
where $\vert n,k\rangle$ are the Bloch functions, $k$ is the Bloch wave vector, $\epsilon_{n,k}$ is the electronic band energy,  $\hat{v}_i$ is the velocity operator in the $i$ direction and $\hat{j}^z_x$=$\frac{1}{2}\{\hat{v}_x,\hat{s}_z\}$ is the spin current operator.

Vibrational frequencies have been calculated within the perturbation theory method as implemented in VASP using the {\tt phonopy} code \cite{phonopy} to obtain the force constants and vibrational frequencies. In this case, we have used a 2$\times$2$\times$2 supercell with a 8$\times$8$\times$8  {\bf q}-mesh to guarantee convergence up to 0.1 cm$^{-1}$.   

\section{Conclusions}

Band inversion in topological insulators induced by hybridization of the energy bands produce quasi-nodal spheres.
The orbital type change on the bands occurs on the quasi-nodal sphere and the matrix information
on how this change is happening encodes the strong Fu-Kane-Mele invariant.
The signal of the SHC is localized where the orbital type changes, and therefore the signal is strong close
 to the Fermi level, where hybridization occurs.

 The appearance of quasi-nodal spheres on materials with band inversion due to hybridization is therefore enforced.
 The change of orbital type must occur in a 2-dimensional sphere, otherwise the Fu-Kane-Mele invariant is forced to be trivial.  

The band inversion on YH$_3$ and CaTe is confirmed and the SHC is calculated in both materials. The
signal is localized on the quasi-nodal sphere and, by removing the Dirac point of CaTe by strain, it is shown that
in both cases the signal of the SHC is localized on the quasi-nodal spheres where the orbital type changes.
This result is expected to hold in materials with a similar type of band inversion as the trihydrides (XH$_3$, X= Y,Gd,Ho,Tb,Nd)
and Si$_3$N$_2$.

%*** Acknowledgement ***
\section{Acknowledgments}
The first author gratefully acknowledges the computing time granted on the supercomputer Mogon at Johannes Gutenberg University Mainz (hpc.uni-mainz.de). 
The second author thanks the Royal Society of the UK for an International Collaboration Award and the Science and Technology Facilities Council of the UK for computing time on SCARF.  
The third author acknowledges the support of CONACYT through project CB-2017-2018-A1-S-30345-F-3125. The first and the third authors thank the continuous support of the Alexander Von Humboldt Foundation, Germany.

%*** Bibliography ***
%\newpage

\bibliographystyle{apsrev}
\bibliography{bibliography}

\end{document}